\documentclass[12pt,journal,draftcls,a4paper,onecolumn]{IEEEtran} 
\usepackage{dsfont}
\usepackage{subeqnarray}
\usepackage{cleveref}
\usepackage{amssymb}
\usepackage{amsfonts}
\usepackage{amsmath}
\usepackage{lipsum}
\usepackage{graphicx}
\usepackage{amsmath}
\usepackage[all,poly]{xy}
\usepackage{multirow}
\usepackage{algorithm}
\usepackage{algorithmic}
\usepackage{color}
\usepackage[noadjust]{cite}

\title{Collaborative sparse regression using spatially correlated supports -- Application to hyperspectral unmixing}
\author{Yoann Altmann, Marcelo Pereyra and Jose Bioucas Dias
\thanks{Supported in part by the Direction G\'en\'erale de l'armement, French Ministry of Defence, the EPSRC via grant EP/J015180/1, the SuSTaIN program - EPSRC grant EP/D063485/1 - at the Department of Mathematics, University of Bristol, by the Portuguese Science and Technology Foundation, Projects UID/EEA/50008/2013 and PTDC/EEI-PRO/1470/2012. M. Pereyra holds a Marie Curie Intra-European Fellowship for Career Development.}
\thanks{Y. Altmann is with the School of Engineering and Physical Sciences, Heriot-Watt University, Edinburgh
U.K. (email: Y.Altmann@hw.ac.uk).}
\thanks{M. Pereyra is with the School of Mathematics of the University of Bristol, Bristol
U.K. (email: marcelo.pereyra@bristol.ac.uk).}
\thanks{J.M. Bioucas-Dias is with the Instituto de Telecomunica\c{c}\~oes and Instituto Superior T\'ecnico, Universidade de Lisboa, Portugal (email: bioucas@lx.it.pt).}}

\newcommand{\bs}{\boldsymbol{s}}

\newcommand{\bx}{\boldsymbol{x}}

\newcommand{\bz}{\boldsymbol{z}}

\newcommand{\bR}{\boldsymbol{R}}

\newcommand{\bbeta}{\boldsymbol{\beta}}

\newcommand{\bsigma}{\boldsymbol{\sigma}}
\newcommand{\bSigma}{\boldsymbol{\Sigma}}

\newcommand{\argmax}{\operatornamewithlimits{argmax}}

\graphicspath{{figures/}}

\def\bfs{{\mathbf{s}}}

\def\bfx{{\mathbf{x}}}

\def\bfz{{\mathbf{z}}}

\def\bfD{{\mathbf{D}}}

\def\bfS{{\mathbf{S}}}

\def\bfX{{\mathbf{X}}}

\def\bfZ{{\mathbf{Z}}}

\def\bbR{{\mathbb{R}}}

\newcommand{\Vpix}[1]{\mathbf{y}_{#1}}

\newcommand{\MATpix}{\mathbf{Y}}

\newcommand{\nbpix}{N}

\newcommand{\nbband}{L}

\newcommand{\nbmat}{R}

\newcommand{\MATmat}{{\mathbf M}}
\newcommand{\Vmat}[1]{{\mathbf m}_{#1}}

\newcommand{\MATabond}{{\mathbf A}}

\newcommand{\abond}[2]{{a}_{#1,#2}}
\newcommand{\Vabond}[1]{{\boldsymbol{a}}_{#1}}

\newcommand{\Vabonds}{{\boldsymbol{a}}}

\newcommand{\Vnoise}[1]{{\mathbf e}_{#1}}

\newcommand{\transp}{^T}

\newcommand{\Ndistr}[1]{\mathcal{N}\left(#1\right)}

\newcommand{\norm}[1]{\left\|#1\right\|}

\newcommand{\Vzero}{\boldsymbol{0}}

\newcommand{\Indicfun}[2]{\textbf{1}_{#1}\left(#2\right)}

\newenvironment{algogo}[1]{
\smallskip
\noindent \hrule\vspace{0.2\baselineskip} \hrule
\begin{small}
\refstepcounter{algo} \center{\bf \textsc{Algorithm \thealgo}}
\\{\center{\bf #1}}
\smallskip
\flushleft
 } {
\end{small}
\smallskip
\hrule\vspace{0.2\baselineskip} \hrule
\smallskip
}

\newcounter{algo}
\renewcommand{\thealgo}{\arabic{algo}}

\begin{document}
\maketitle

\begin{abstract}
This paper presents a new Bayesian collaborative sparse regression method for linear unmixing of hyperspectral images. Our contribution is twofold; first, we propose a new Bayesian model for structured sparse regression in which the supports of the sparse abundance vectors are \emph{a priori} spatially correlated across pixels (\emph{i.e.}, materials are spatially organised rather than randomly distributed at a pixel level). This prior information is encoded in the model through a truncated multivariate Ising Markov random field, which also
takes into consideration the facts that pixels cannot be empty (i.e, there is at least one material present in each pixel), and that different materials may exhibit different degrees of spatial regularity. Secondly, we propose an advanced Markov chain Monte Carlo algorithm to estimate the posterior probabilities that materials are present or absent in each pixel, and, conditionally to the maximum marginal \emph{a posteriori} configuration of the support, compute the MMSE estimates of the abundance vectors. A remarkable property of this algorithm is that it self-adjusts the values of the parameters of the Markov random field, thus relieving practitioners from setting regularisation parameters by cross-validation. The performance of the proposed methodology is finally demonstrated through a series of experiments with synthetic and real data and comparisons with other algorithms from the literature.
\end{abstract}


\section{Introduction}
\label{sec:introduction}
Spectral unmixing (SU) of hyperspectral images is a challenging problem that has received a lot of attention over the last few years \cite{Bioucas2012,Bioucas2013,Dobigeon_IEEE_SP_Mag_2014}. It consists in identifying the materials (endmembers) present in an image and simultaneously quantifying their fractions or proportions within each pixel (abundances). This source separation problem has been widely studied for applications where pixel reflectances are linear combinations of pure component spectra \cite{Craig1994,Heinz2001,Eches2010a,Miao2007,Yang2011}. The typical SU processing pipeline is then decomposed into three main estimation steps:  the estimation of the number of different materials present the image, the estimation or extraction of their spectral signatures, and finally the quantification of their abundances.

Abundance estimation is known to be a challenging problem, in particular in scenarios involving materials with similar spectral signatures. In these cases, exploiting prior knowledge about the problem can improve estimation performance dramatically. A particularly important form of prior knowledge is that the number of materials within each pixel is typically much smaller than the total number of materials present in the scene (this property is accentuated in modern images that are acquired with high spatial resolution sensors). In other words, the abundance vectors are generally sparse. Sparsity also arises naturally when SU is performed with a dictionary or library containing the spectral signatures of a large number materials, some of which are possibly present in the scene \cite{Iordache2011}. 

Once sparsity is taken into consideration, SU can be conveniently formulated as a sparse regression problem whose objective is to jointly identify the materials within each pixel and to quantify their abundance. This regression problem is often solved by penalised maximum likelihood estimation, which can be efficiently computed with state-of-the-art optimisation algorithms (typically an $\ell_1$ penalty is used to promote sparse solutions) \cite{Bioucas2010whispers}. Recently, Iordache \emph{et al.} \cite{Iordache2014a} proposed a collaborative sparse regression technique (CLSunSAL) based on an $\ell_{2,1}$ penalty function that enforces group-sparsity for the abundances of each material. This method was further improved in \cite{Iordache2014b} by introducing a pre-processing step that identifies the elements from the spectral library that are more likely present in the image. The resulting MUSIC-CSR algorithm solves a sparse regression problem that is collaborative in the sense that all the image pixels are used to identify the active endmembers. Sparse regression for SU can also be successfully performed within the Bayesian framework. For example, Dobigeon \emph{et al.} \cite{Dobigeon2008} and Eches \emph{et al.} \cite{Eches2010} propose Bayesian models and Monte Carlo algorithms to identify the active endmembers in an HSI from a spectral library while ensuring that the abundances of absent endmembers are zero. Note that library-based methods are not the only strategy to address the absence of pure pixels (see \cite{Craig1994,Miao2007a,Bioucas2009whispers,Nascimento2012} for more details).

It is widely acknowledged that collaborative sparse regression methods can produce very accurate SU results. Collaboration is key because it improves the estimation of the support of the sparse abundance vectors, thus reducing significantly the number of unknowns. However, most existing collaborative techniques only exploit global information and therefore can only seek to determine if an endmember is present/absent in the entire image (that is, can only estimate the union of the supports of all the abundance vectors in the image). As a result, global collaborative techniques may overestimate significantly the support of the actual abundance vector of each pixel. 

This paper presents a new collaborative sparse regression technique that exploits the local spatial correlations in the image to accurately detect the endmembers that are active/inactive in each pixel. Precisely, we present a Bayesian model that simultaneously promotes sparsity on the abundance vectors, and spatial correlation between the abundance vectors' supports (\emph{i.e.}, non-zero elements), modelling the spatial presence and absence patterns of materials in the scene. This approach differs significantly from the strategies adopted in the previous works \cite{Eches2011,Iordache2012,Eches2013,Chen2014,Altmann2014a}, where spatial correlations are introduced by regularising the abundance values or the nonlinear effects occurring in the image. The latter strategies  perform well in hyperspectral images with low spatial resolution, and composed mainly of homogenous regions, but are inadequate for high resolution images and for images involving complex scenes, small targets, and textures or fluctuations in the abundances.
By operating directly at the level of the abundance vectors' support, our model is able to capture spatial correlations in a more subtle manner and produce accurate estimation results in challenging scenes involving, for example, crops (textured scenes) and isolated trees.

The remainder of the paper is organized as follows. Section \ref{sec:Problem} recalls the classical linear mixing model for SU and presents the proposed Bayesian model for sparse regression. In Section \ref{sec:sampler} we propose an original Monte Carlo method to perform Bayesian inference in this model and perform SU. The proposed methodology is demonstrated on synthetic and real HSI in in Sections \ref{sec:simu_synth} and \ref{sec:simu_real}. Conclusions are finally reported in Section
\ref{sec:conclusion}.

\section{Problem statement}
\label{sec:Problem}
Consider a hyperspectral image $\MATpix \in \bR^{\nbband \times \nbpix}$, 
where $\nbband$ is the number of spectral bands considered and $N=N_{\textrm{row}}\times N_{\textrm{col}}$ corresponds to the total number of pixels. Under the linear mixing assumption, each image pixel $\Vpix{n}=\left[\MATpix_{1,n}, \ldots, \MATpix_{L,n} \right]\transp \in \bR^\nbband$ can be expressed as a linear combination of $\nbmat$ known spectral signatures $\Vmat{1},\ldots,\Vmat{\nbmat}$ corrupted by zero-mean Gaussian noise with diagonal covariance $\bSigma_0$, that is,
\begin{eqnarray} \label{eq:model}
\Vpix{n} = \sum_{r=1}^{\nbmat}{\abond{r}{n}\Vmat{r}} + \Vnoise{n}, \quad \Vnoise{n} \sim \Ndistr{\Vzero,\bSigma_0}
\end{eqnarray}
where $\abond{r}{n}$ is the mixing coefficient associated with the $r$th endmember in the $n$th pixel and $\bSigma_0=\textrm{diag}(\bsigma^2)$ with $\bsigma^2=[\sigma_1^2,\ldots,\sigma_{L}^2]\transp$.
By setting $\MATmat=[\Vmat{1},\ldots,\Vmat{\nbmat}]$ and $\Vabond{n}=[\abond{1}{n},\ldots,\abond{R}{n}]\transp$, Eq. \eqref{eq:model} can be conveniently expressed in matrix notation as $\Vpix{n} = \MATmat\Vabond{n} + \Vnoise{n}$. 

This paper considers the supervised spectral unmixing problem of the hyperspectral image $\MATpix$, \emph{i.e.},  the estimation of the $R \times N$ abundance matrix $\MATabond=[\Vabond{1},\ldots,\Vabond{N}]$. More precisely, we consider Bayesian methods for estimating $\MATabond$ given $\MATpix$ (and the endmember matrix $\MATmat$) subject to the following two sets of physical constraints: first, the abundances are non-negative quantities, \emph{i.e.}, $\abond{r}{n} \geq 0 \quad \forall r, n$; second, there is at least one material present in each pixel and therefore $\norm{\Vabond{n}}_0>0$, where $\norm{\cdot}_0$ denotes the $\ell_0$ vector pseudo-norm. We also assume that the values of the noise variances $\bsigma^2$ are unknown, though prior knowledge, if available, can be easily integrated into the model.

In a manner akin to \cite{Newstadt2014ssp}, we model explicitly the sparsity of $\MATabond$ by using the decomposition
\begin{eqnarray}
\label{eq:abond_decomp}
\MATabond = \bfZ \odot \bfX
\end{eqnarray}
where $\bfZ \in \left\lbrace 0,1 \right\rbrace^{{ \nbmat \times N}}$ is a matrix of Bernoulli variables that ``labels'' each material as \emph{present} (active) or \emph{absent} (inactive) in each pixel, $\bfX \in \bR^{\nbmat \times N}$ is a matrix with positive entries that (jointly with $\bfZ$) quantifies the abundances, and $\odot$ denotes the Hadamard (term-wise) matrix product. This decomposition is particularly useful in Bayesian sparse regression problems because it allows eliciting separate statistical models for $\Vabond{n}$'s support (through modelling $\bfZ$), and for the values of the positive elements of $\Vabond{n}$ (through $\bfX$).

The next section presents a Bayesian model for estimating $\bfZ$ and $\bfX$ subject to the physical constraints discussed above. A key aspect of this model is that it will capture the fact that the pixels in which a material is active (or inactive) generally form spatial clusters (\emph{i.e.}, exhibit spatial group sparsity). In difficult unmixing scenarios, exploiting this strong prior information can significantly improve estimation results, as will be shown in this paper.


\section{Bayesian Model}
\label{sec:bayesian_model} 
This section presents the proposed the hierarchical Bayesian model for performing sparse source separation with collaborative supports. This model is defined by specifying  the likelihood and the prior distribution of the parameters of interested $\bfZ$ and $\bfX$, as well as for the other unknown parameters in the model (e.g. $\bsigma^2$) that will be subsequently removed by marginalisation (\emph{i.e.}, integrated out of the model's joint posterior distribution).

\subsection{Likelihood}
From the observation model \eqref{eq:model} and the parametrisation of $\MATabond$ described in \eqref{eq:abond_decomp}, the likelihood of the image $\MATpix$ given the unknown parameters is
\begin{eqnarray}
f(\MATpix|\bfZ,\bfX,\MATmat,\bsigma^2) &= &\prod_{n} f(\Vpix{n}|\bz_{n}, \bx_{n},\MATmat,\bsigma^2 )\\
&= &\prod_{n} p_{\mathcal{N}} (\Vpix{n}|\MATmat(\bz_{n} \odot \bx_{n}),\bSigma_0)
\end{eqnarray}
where $\bz_{n}$ (resp. $\bx_{n}$)is the $n$th column of $\bfZ$ (resp. $\bfX$) and $p_{\mathcal{N}}(\Vpix{n}|\MATmat(\bz_{n} \odot \bx_{n}),\bSigma_0)$ is the probability density function of a multivariate Gaussian vector with mean vector $\MATmat(\bz_{n} \odot \bx_{n})$ and diagonal covariance matrix $\bSigma_0$.

\subsection{Prior distribution of $\bfZ$}
As explained previously, a key aspect of the proposed Bayesian model is to take into account the fact that the pixels in which a given material is present or absent typically form spatial groups or clusters (as opposed to being randomly distributed in space). From a modelling viewpoint, this can be represented by correlating the Bernoulli variables or labels $\bz_{r,n}$ across the spatial dimension indexed by $n$. This can be achieved, for example, by stating that if a certain material is present (or absent) in a given pixel, this increases the probability of its presence (or absence) in neighbouring pixels. Taking into account that each material can exhibit its own spatial configuration, and the constraint that there must be at least one material present in each pixel, we propose to assign $\bfZ$ the following \emph{truncated multivariate Ising Markov random field} prior
\begin{eqnarray}\label{eq:MRF}
f(\bfZ | \bbeta) =\dfrac{\psi(\bfZ)}{C(\bbeta)} \exp{\left[ \sum_{r=1}^R \beta_r  \phi_r (\bfZ)\right]}
\end{eqnarray}
with $\bbeta = \{\beta_1,\ldots,\beta_R\}$,
\begin{eqnarray}
\phi_r (\bfZ) &=&  \sum_{n} \sum_{n' \in \mathcal{V}(n)} \delta(z_{r,n} -
z_{r,n'}),\\
\psi(\bfZ) &=& \prod_{n} \underset{r}{\textrm{max}}(z_{r,n}),\label{eq:psi}\\
C(\bbeta) &=& \sum_{\bfZ} \psi(\bfZ) \exp{\left[ \sum_{r=1}^R \beta_r  \phi_r (\bfZ)\right]},
\end{eqnarray}
and where $\delta(\cdot)$ is the Kronecker delta function and $\mathcal{V}(n)$ denotes the set of neighbours of pixel $(n)$ (in this paper we have used the 8-pixel neighbourhood). The hyper-parameters 
$\beta_1,\ldots,\beta_R$ act as regularization parameters that control the degree of spatial smoothness or regularity associated with each endmember, accounting for the fact that different materials may exhibit different spatial distributions.

To gain intuition about the proposed prior, we note that setting $\bbeta = \boldsymbol{0}$ in \eqref{eq:MRF} and considering the prior \eqref{eq:prior_x} for the matrix $\bfX$ leads to a Bernoulli-Gaussian type prior for the abundances which is closely related to the $\ell_0$-$\ell_2$ penalty often used for sparse regression. In the general scenarios where $\bbeta \neq \boldsymbol{0}$, Eq. \eqref{eq:MRF} introduces spatial correlations between the components of $\bfZ$ and, together with \eqref{eq:prior_x}, leads to an $\ell_0$-$\ell_2$-type penalty promoting spatial group sparsity for the abundances. Prior \eqref{eq:MRF} can also be understood as a collaborative prior, in the sense that, by assigning higher probabilities to configurations in which the supports of the abundance vectors $\Vabond{n}$ are spatially correlated, it pools or shares information between the sparse regressions that take place at each pixel. Crucially, by capturing the spatial correlations that occur naturally in hyperspectral images, this prior can improve significantly estimation results.
%

\subsection{Prior distribution of $\bfX$}
We assign the elements of $\bfX$ the following hierarchical prior distribution
\begin{eqnarray}\label{eq:prior_x}
x_{r,n} | s_r^2 &\sim &\mathcal{N}_{\bbR^+}\left(0,s_r^{2} \right)\\
s_r^2|\gamma,\nu &\sim &\mathcal{IG}(\gamma,\nu)
\end{eqnarray}
parametrised by some fixed hyper-parameters $(\gamma,\nu)$, and where we note that the prior on $x_{r,n} | s_r^2$ is truncated to $\bbR^+$ to reflect the positivity of $x_{r,n}$. This prior is very flexible and can be adjusted to represent a wide variety of prior beliefs (in all our experiments we used $\gamma = 2.1$ and $\nu = 1.1$, corresponding to a weakly informative prior for $s_r^2$ with $80\%$ of its mass in $[0,1]$.)

An important advantage of the hierarchical prior \eqref{eq:prior_x} is its natural capacity to encode prior dependences between the abundances. We expect the abundance coefficients associated with the same material to exhibit correlations, in particular in terms of their scale. This belief is encoded in \eqref{eq:prior_x} by defining a common parameter $s_r^2$ for each material or endmember $\Vmat{r}$, which is shared by all the abundances related to that material. Therefore, the hierarchical structure of \eqref{eq:prior_x} operates as a global pooling mechanism that shares information across the rows of $\MATabond$ (i.e, the abundance coefficients associated to the $r$th material) to improve estimation accuracy.

Assuming an exchangeable structure where the abundances are prior independent given the hidden variables $\bfs^2=[s_1^{2},\ldots,s_R^{2}]\transp$, we obtain the following following joint prior for $\bfX, \bfs^2$ 
\begin{eqnarray}
\label{eq:joint_abund_prior}
f(\bfX, \bfs^2) = f(\bfX| \bfs^2)f(\bfs^2),
\end{eqnarray}
with $f(\bfX| \bfs^2)  = \prod_{r,n} f(x_{r,n}|s_r^{2})$ and $f(\bfs^2) = \prod_{r} f(s_r^2|\gamma,\nu)$. Also notice that by using the hierarchical structure \eqref{eq:prior_x} we obtain conjugate priors and hyper-priors for $x_{r,n}$ and $s_r^{2}$; this leads to inference algorithms with significantly better tractability and computational efficiency, which is crucial given the high dimensionality of $\bfX$.

Finally, notice that in the proposed model the spatial dependences in the abundance maps are encoded at the level of the abundance supports (though the prior on $\bfZ$), and not directly through the values of the abundances $\MATabond = \bfZ \odot \bfX$ as it is the case in some previous models (see for example \cite{Iordache2012}). The motivation for this modelling choice is that modern high-resolution hyperspectral images often exhibit textures and fine detail that are not well described by models that promote smooth or piecewise-constant abundances maps. The experiments reported in this work show that the two approaches to modelling spatial correlation have complementary strengths and weaknesses, suggesting that to further improve estimation results future models should consider both levels of spatial dependences simultaneously.


\subsection{Prior distribution of the noise variances $\bsigma^2$}
In this paper we consider that there is no significant prior knowledge available about the values of the noise variances and assign each $\sigma_{\ell}^2$ its non-informative Jeffreys prior \cite{Bernardo94}
\begin{eqnarray}
\label{eq:sigma2r}
f(\sigma_{\ell}^2) \propto \dfrac{1}{\sigma_{\ell}^2} \Indicfun{\bbR^+}{\sigma_{\ell}^2},\quad \ell=1,\ldots,L.
\end{eqnarray}
Note that in scenarios where prior knowledge about $\bsigma^2$ is available, this can be easily integrated into the model by replacing \eqref{eq:sigma2r} with an inverse gamma conjugate priors with hyper-parameter values reflecting this prior knowledge.


\subsection{Regularisation parameter $\bbeta$}
A main advantage of Bayesian methods is that they allow estimating the appropriate amount of regularisation from data, thus freeing practitioners from the difficulty of setting regularisation parameters by cross-validation. Indeed, there are several Bayesian strategies for selecting the value of the regularisation parameter $\bbeta$ in a fully automatic manner (see \cite{Pereyra2013ip} for a recent detailed survey on this topic). In this paper we use the empirical Bayes technique recently proposed in \cite{Pereyra2014ssp}, where the value of $\bbeta$ is estimated by maximum marginal likelihood.

\subsection{Joint posterior distribution of $\bfZ, \bfX, \bsigma^2, \bfs^2$}
The structure of the proposed hierarchical Bayesian model is summarised in the directed acyclic graph (DAG) depicted below in Fig. \ref{fig:DAG}.
\begin{figure}[h!]
\centerline{ \xymatrix{
  & *+<0.05in>+[F-]+{\gamma} \ar@/^/[rd]& *+<0.05in>+[F-]+{\nu} \ar@/^/[d]&  \\
  & \bbeta \ar@/^/[d]   & \bs^2 \ar@/^/[d] & \\
  *+<0.05in>+[F-]+{\MATmat} \ar@/_/[rd]    & \bfZ \ar@/_/[d] &   \bfX \ar@/^/[ld]& \bsigma^2 \ar@/^/[lld]  \\
   & \MATpix &   & }
} \caption{Directed acyclic graph (DAG) of the proposed hierarchical Bayesian model (parameters with fixed values are represented using black boxes).} \label{fig:DAG}
\end{figure}
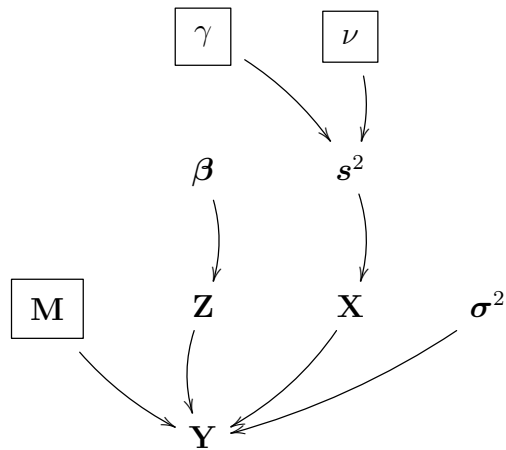

Using Bayes' theorem, and taking into account the conditional independences of the Bayesian model (see Fig. \ref{fig:DAG}), the joint posterior distribution of $\bfZ, \bfX, \bsigma^2$ and $\bfs^2$ given the observations $\MATpix$, the library of spectral signatures $\MATmat$ and the model's fixed parameters $\bbeta, \gamma$ and $\nu$, is given by
\begin{eqnarray}
\label{eq:posterior}
f(\bfZ,\bfX,\bsigma^2,\bfs^2| \MATpix, \MATmat, \bbeta, \gamma, \nu )\propto f(\MATpix|\bfZ,\bfX,\MATmat,\bsigma^2) f(\bfZ|\bbeta) f(\bfX | \bfs^2) f(\bfs^2|\gamma,\nu) f(\bsigma^2).
\end{eqnarray}

The following section presents a Monte Carlo algorithm to perform Bayesian inference using the proposed model.

\section{Bayesian inference using Gibbs sampling}
\label{sec:sampler} 
The Bayesian model defined in Section \ref{sec:bayesian_model} specifies a joint posterior distribution for the unknown parameters $\bfZ, \bfX, \bsigma^2, \bfs^2$ given the fixed quantities $\MATpix, \MATmat, \gamma,\nu$ and the hyper-parameter $\bbeta$ which is unknown but represented as a deterministic parameter (whose value will be tuned during the inference procedure). According to the Bayesian paradigm, this posterior distribution fully describes the information about the unknowns that is provided by the data and by the prior knowledge available. However, for unmixing applications it is necessary to summarise this posterior distribution in the form of point estimates; that is, to assign specific values for the unknown quantities of interest (in our problem the abundance vectors). Here we consider the following coupled Bayesian estimators that are particularly suitable for sparse regression problems: the marginal maximum \emph{a posteriori} (MMAP) estimator \cite{Robert2007,Doucet2002} for the support of the abundance vectors or ``presence maps''
\begin{eqnarray}\label{zEstimator}
z^{MMAP}_{r,n} = \argmax_{z_{r,n} \in \{0,1\}} f (z_{r,n} | \MATpix, \MATmat, \bbeta, \gamma,\nu),
\end{eqnarray}
and, conditionally on the estimated supports, the minimum mean square error estimator of the abundances
\begin{eqnarray}\label{aEstimator}
a^{MMSE}_{r,n} = \textrm{E}\left[ x_{r,n} | z_{r,n} = \hat{z}^{MMAP}_{r,n},  \MATpix, \MATmat, \bbeta, \gamma,\nu \right],
\end{eqnarray}
where 
\begin{eqnarray*}
f (z_{r,n}|\MATpix, \MATmat, \bbeta, \gamma,\nu)= \int f(\bfZ,\bfX,\bsigma^2,\bfs^2| \MATpix, \MATmat, \bbeta, \gamma, \nu ) \textrm{d}\bfZ_{\backslash z_{r,n}} \textrm{d}\bfX, \textrm{d}\bsigma^2 \textrm{d}\bfs,
\end{eqnarray*}
with the matrix $\bfZ_{\backslash z_{r,n}}$ containing the remaining elements of $\bfZ$ once $z_{r,n}$ has been removed and where $\textrm{E}\left[\cdot\right]$ denotes the expectation with respect to the conditional marginal density
\begin{eqnarray*}
f (x_{r,n} | z_{r,n}, \MATpix, \MATmat, \bbeta, \gamma,\nu) = \frac{\int f(\bfZ,\bfX,\bsigma^2,\bfs^2| \MATpix, \MATmat, \bbeta, \gamma, \nu ) \textrm{d}\bfZ_{\backslash z_{r,n}} \textrm{d}\bfX_{\backslash z_{r,n}}, \textrm{d}\bsigma^2 \textrm{d}\bfs}{f (z_{r,n} | \MATpix, \MATmat, \bbeta, \gamma,\nu)}.
\end{eqnarray*}
Note that the abundance estimator \eqref{aEstimator} is sparse by construction (\emph{i.e.}, $\textrm{E}\left[ x_{r,n} | z_{r,n} = 0,  \MATpix, \MATmat, \bbeta, \gamma,\nu \right] = 0$) and that, by marginalising out the other unknowns, it automatically takes into account the uncertainty about $\sigma^2$ and $\bfs^2$. 

The choice of specific Bayesian estimators to summarise the posterior distribution is a decision-theoretic problem that depends on the model and the application considered \cite{Robert2007}. In the model described in this work, the two quantities of interest $\boldsymbol{\textrm{Z}}$ and $\boldsymbol{\textrm{X}}$ are very different in nature, and as a result their posterior information is best summarised with different estimators. The labels $\boldsymbol{\textrm{Z}}$ are binary variables that describe a quantitative aspect of the model; that is, they parametrise the hypotheses that materials are present or absent in each pixel. Selecting the hypotheses with highest posterior probability leads to marginal MAP estimation \cite{Robert2007}, which in this case operates as a model-selection tool. Moreover, the conditional posterior distribution of $\boldsymbol{\textrm{X}}|\boldsymbol{\textrm{Z}}$ represents the uncertainty regarding the abundance values for a specific model or configuration (in particular the one identified by marginal MAP estimation). To summarise this posterior distribution we use the MMSE estimator because it is optimal with respect to any quadratic loss function, and approximately optimal with respect to any convex loss function \cite{Robert2007}. Furthermore, in cases where there is strong prior knowledge justifying the use of a sum-to-one constraint on the abundances, this information can be incorporated to the inferences by using an MMSE estimator constrained to the simplex. However, we have observed that this property does not hold on our images, in part because of the effects of mild shadows in the scene.

Computing \eqref{zEstimator} and \eqref{aEstimator} is challenging because it requires having access to the univariate marginal densities of $z_{r,n}$ and the joint marginal densities of $(x_{r,n}, z_{r,n})$, which in turn require computing the posterior \eqref{eq:posterior} and integrating it over a very high-dimensional space. Fortunately  these estimators can be efficiently approximated with arbitrarily large accuracy by Monte Carlo integration. Precisely, it is possible to compute \eqref{zEstimator} and \eqref{aEstimator}  by first using a Markov Chain Monte Carlo (MCMC) computational method to generate samples asymptotically distributed according to \eqref{eq:posterior}, and subsequently using these samples to approximate the required marginal probabilities and expectations. 

Here we propose a Gibbs sampler to simulate samples from \eqref{eq:posterior}, as this type of MCMC method is particularly suitable for models involving hidden Markov random fields \cite[Chap. 10]{Robert2004}. The output of this algorithm are two Markov chains of $N_{\textrm{MC}}$ samples $\{\bfX^{(1)},\ldots, \bfX^{(N_{\textrm{MC}})}\}$ and $\{\bfZ^{(1)},\ldots, \bfZ^{(N_{\textrm{MC}})}\}$ that are asymptotically distributed according to the posterior distribution $f(\bfZ,\bfX| \MATpix, \MATmat, \bbeta, \gamma, \nu )$. The first $N_{\textrm{bi}}$ samples of these chains correspond to the so-called \emph{burn-in} transient period and should be discarded (the length of this period can be assessed visually from the chain plots or by computing convergence tests). The remaining  $N_{\textrm{MC}} - N_{\textrm{bi}}$ of each chain are used to approximate the Bayesian estimators \eqref{zEstimator} and \eqref{aEstimator} as follows
\begin{eqnarray}\label{zEstimator2}
\hat{z}^{MMAP}_{r,n} = \argmax_{u = \{0,1\}} \sum_{t = N_{\textrm{bi}} + 1}^{N_{\textrm{MC}}} \delta\left(z^{(t)}_{r,n} - u\right),
\end{eqnarray}
and 
\begin{eqnarray}\label{aEstimator2}
\hat{a}^{MMSE}_{r,n} = \frac{\sum_{t = N_{\textrm{bi}} + 1}^{N_{\textrm{MC}}} x^{(t)}_{r,n} \delta\left(z^{(t)}_{r,n} - \hat{z}^{MMAP}\right)}{\sum_{t = N_{\textrm{bi}} + 1}^{N_{\textrm{MC}}} \delta\left(z^{(t)}_{r,n} - \hat{z}^{MMAP}_{r,n}\right)}.
\end{eqnarray}
Note that \eqref{zEstimator2} and \eqref{aEstimator2} converge to the true Bayesian estimators \eqref{zEstimator} and \eqref{aEstimator} as $N_{\textrm{MC}} \rightarrow \infty$. The remainder of this sections provides details about the main steps of the proposed Gibbs sampler, termed {\em Collaborative sparse Unmixing} (CSU) and summarised in Algo. \ref{algo:algo1} below. Note that for clarity the dependence of all distributions on the known fixed quantities $\MATmat, \gamma, \nu$ and $\bbeta$ is omitted.

\begin{algogo}{Collaborative sparse Unmixing (CSU)}
     \label{algo:algo1}
     \begin{algorithmic}[1]
        \STATE \underline{Fixed input parameters:} $\MATmat,K,\gamma,\nu$, number of burn-in iterations $N_{\textrm{bi}}$, total number of iterations $N_{\textrm{MC}}$
				\STATE \underline{Initialization ($t=0$)}
        \begin{itemize}
        \item Set $\bfX^{(0)},\bfZ^{(0)},\bsigma^{2(0)},\bs^{2(0)}, \bbeta^{(0)}$
        \end{itemize}
        \STATE \underline{Iterations ($1 \leq t \leq N_{\textrm{MC}}$)}
				\STATE Set $\bfZ^{*}=\bfZ^{(t-1)}$
				\FOR{$n=1:N$}
        \STATE Sample $\bfz_{n}^* \sim f(\bfz_{n}^*|\MATpix,\bfZ_{\backslash \bfz_{n}^*}^*\bfX^{(t-1)},\bsigma^{2(t-1)},\bs^{2(t-1)})$ in \eqref{eq:post_z}
				\ENDFOR
				\STATE Set $\bfZ^{(t)}=\bfZ^{*}$
        \STATE Sample $\bfX^{(t)} \sim f(\bfX|\MATpix,\bfZ^{(t)},\bsigma^{2(t-1)},\bs^{2(t-1)},)$ in \eqref{eq:joint_post_abond}
        \STATE Sample $\bsigma^{2(t)} \sim f(\bsigma^{2}|\MATpix,\bfX^{(t)},\bfZ^{(t)},\bs^{2(t-1)},)$ in \eqref{eq:post_sigma2}
				\STATE Sample $\bs^{2(t)} \sim f(\bs^{2}|\MATpix,\bfX^{(t)},\bfZ^{(t)},\bsigma^{2(t)})$ in \eqref{eq:post_bfs}
	\STATE Update $\bbeta^{(t)} \leftarrow \bbeta^{(t-1)}$ using \cite{Pereyra2014ssp}.
        \STATE Set $t = t+1$.
\end{algorithmic}
\end{algogo}

\subsection{Sampling the label matrix $\bfZ$}
The label matrix $\bfZ$ is updated pixel-wise by iteratively simulating from the distribution of the labels at each pixel given the other pixels. Precisely, the distribution of the vector $\bfz_{n}$ given the matrix $\bfZ_{\backslash \bfz_{n}}$ containing the remaining elements of $\bfZ$ once $\bfz_{n}$ is removed is given by
\begin{eqnarray}\label{eq:post_z}
P(\bfz_{n}|\MATpix,\bfZ_{\backslash \bfz_{n}},\bfX,\bsigma^2,\bfs^2)&  \propto &  f(\Vpix{n}|\bfz_{n},\bfx_{n},\bsigma^2) f(\bfz_{n}|\bfZ_{\backslash \bfz_{n}})\\
& \propto &  \exp \left[2 \sum_{r=1}^R \beta_r \sum_{n' \in \mathcal{V}(n)} \delta(z_{r,n} - z_{r,n'})\right] \exp \left[-\dfrac{1}{2\sigma^2} \tilde{\mathbf{y}}_n\transp\bSigma_0^{-1}\tilde{\mathbf{y}}_n\right]
\end{eqnarray}
where $\tilde{\mathbf{y}}_n=\Vpix{n} - \MATmat(\bfx_{n}\odot\bfz_{n})$, for $||\bfz_{n}||_0 > 0$  and $P(\bfz_{n}|\MATpix,\bfZ_{\backslash \bfz_{n}},\bfX,\bsigma^2,\bfs^2) = 0$ otherwise. Algorithmically, this simulation step can be achieved by indexing the $2^R-1$ admissible configurations of $\bfz_{n}$ and then randomly selecting a specific one with probability defined in \eqref{eq:post_z}.

\subsection{Sampling $\bfX$}
The conditional distribution of $\bfX$ given the other unknown parameters can be factorised pixel-wise as a product of $N$ marginal distributions
\begin{eqnarray}
\label{eq:joint_post_abond}
f(\bfX|\MATpix,\bfZ,\bsigma^{2},\bs^{2}) = \prod_{n=1}^N f(\bfx_{n}|\Vpix{n},\bfz_{n},\bsigma^{2},\bs^{2}),
\end{eqnarray}
that can be efficiently sampled independently and in parallel
\begin{eqnarray}
\label{eq:post_abond}
\bfx_{n}|\Vpix{n},\bfz_{n},\bsigma^{2},\bs^{2} \sim \mathcal{N}_{\Omega}(\bar{\bfx}_{n},\bSigma_{n}),
\end{eqnarray}
where $\Omega = \left(\bbR^+\right)^R$ is the positive orthant of $\mathbb{R}^R$ and
\begin{eqnarray}
\bSigma_{n} & = & \left(\bfD_{n}\MATmat\transp\bSigma_0^{-1}\MATmat\bfD_{n} + \bfS^{-1}\right)^{-1}\nonumber\\
\bar{\bfx}_{n} & = & \bSigma_{n} \bfD_{n}\MATmat\transp\bSigma_0^{-1}\Vpix{n},\nonumber
\end{eqnarray}
and where $\bfS=\textrm{diag}(\bfs^2)$ and $\bfD_{n}=\textrm{diag}(\bfz_{n})$ are diagonal matrices with diagonal elements given by $\bfs^2$ and $\bfz_{n}$. For completeness, the derivation of \ref{eq:post_abond} is provided in Appendix. In this paper we use the method \cite{Pakman2012} to simulate efficiently from \eqref{eq:post_abond}.

\subsection{Sampling the noise variances $\bsigma^2$}
It can be easily shown that the noise variances are (conditioned on the other parameters) a posteriori dependent and can thus be updated in a parallel manner. Precisely, the conditional distribution associated with $\sigma_{\ell}^2$ has a simple closed form expression and is given by
\begin{eqnarray}
\label{eq:post_sigma2}
\sigma_{\ell}^{2}|\MATpix,\bfX,\bfZ,\bs^{2} \sim\mathcal{IG}\left(N/2, E_{\sigma_{\ell}} \right),
\end{eqnarray}
with $E_{\sigma_{\ell}} = \norm{\tilde{\mathbf{y}}_{\ell,:}}_2^2/2$ ($\tilde{\mathbf{y}}_{\ell,:}$ denotes the $\ell$th row of the $\nbband \times N$ matrix $\widetilde{\MATpix}=[\tilde{\mathbf{y}}_1,\ldots,\tilde{\mathbf{y}}_N]$).

\subsection{Sampling the hyperparameter vector $\bfs^2$}
The conditional distribution of $\bfs^2$ can be factorised endmember-wise as a product of $R$ independent marginal distributions
\begin{eqnarray}
\label{eq:post_bfs}
f(\bs^{2}|\MATpix,\bfX,\bfZ,\bsigma^{2})=\prod_{r=1}^R f(s_r^{2}|\bfX)
\end{eqnarray}
that can be easily sampled independently and in a parallel manner
\begin{eqnarray}
s_r^{2}|\bfX \sim \mathcal{IG}\left(\dfrac{N}{2} + \gamma, \dfrac{\sum_{n}\bfx_{r,n}^2}{2} + \nu \right).
\end{eqnarray}

\section{Validation with synthetic data}
\label{sec:simu_synth}
This section demonstrates the proposed methodology on a series of experiments conducted using synthetic data. An applications to a real hyperspectral image is reported in 
Section \ref{sec:simu_real}.

\begin{figure}[h!]
  \centering
  \includegraphics[width=\columnwidth]{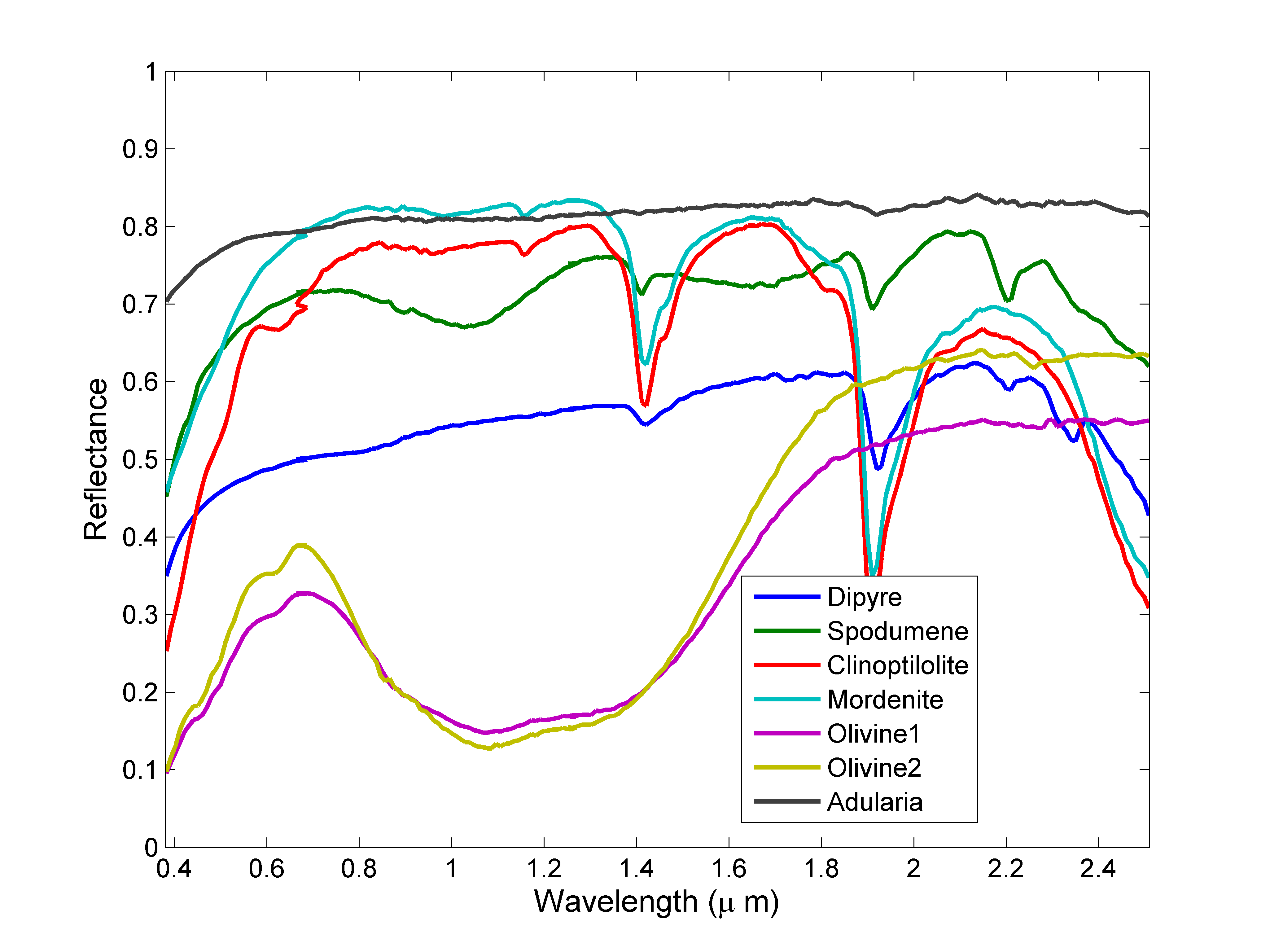}
  \caption{Seven endmembers from the USGS spectral library.}
  \label{fig:endmembers_synth}
\end{figure}
\subsection{Data sets}
\label{subsec:synth_data}
\begin{figure}[h!]
  \centering
	  \includegraphics[width=\columnwidth]{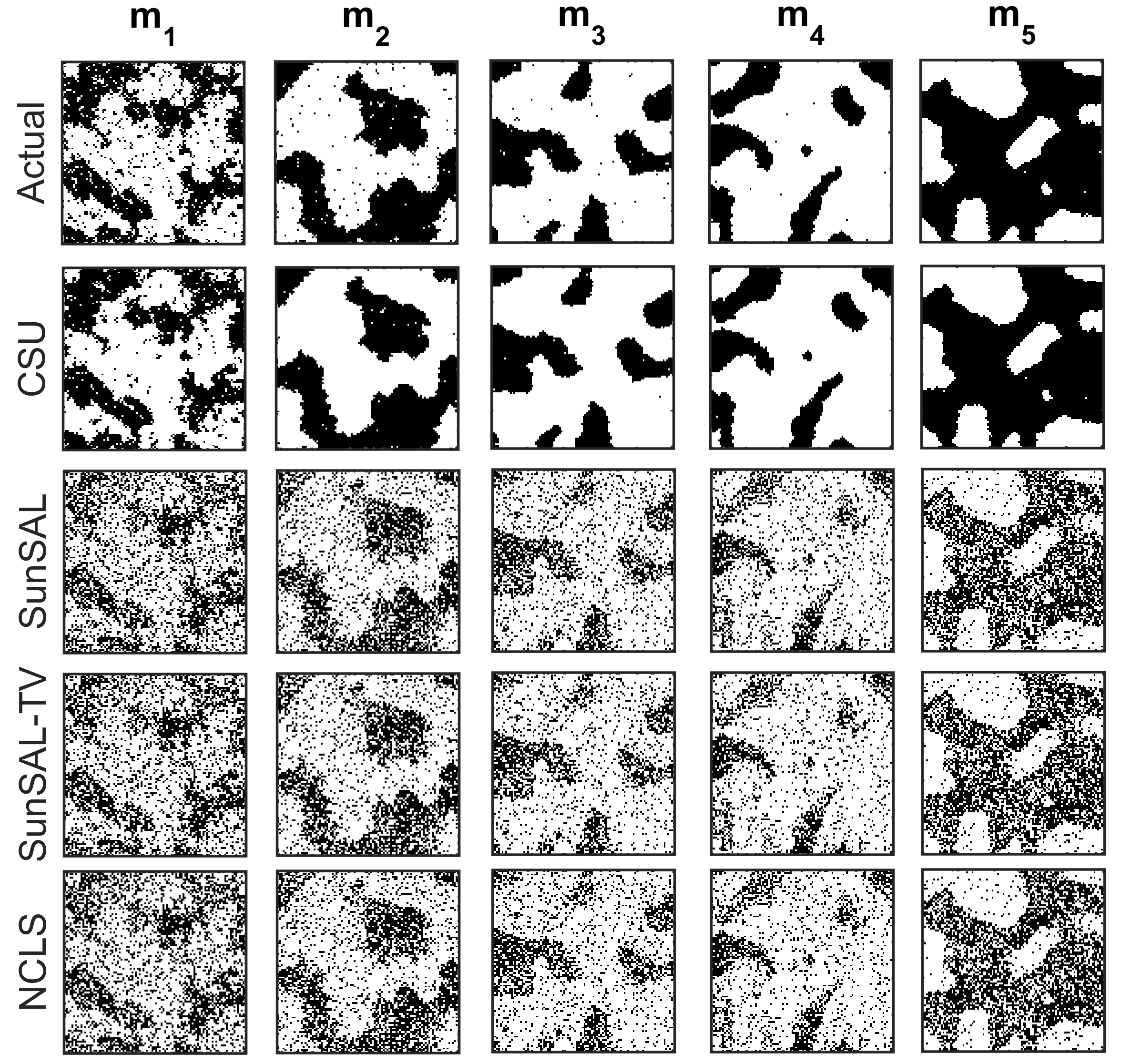}
  \caption{Top row: Active support maps of the $R_0=5$ endmembers associated with $I_1$ and $I_2$. Active support maps estimated with the proposed CSU algorithm, SunSAL, SunSAL-TV, and NCLS ($\rho=0.01$) for $I_2$ (top to bottom). White (resp. black) pixels correspond to regions where a component is present (resp. absent).}
  \label{fig:Potts_synth_R5}
\end{figure}

The performance of the proposed collaborative sparse unmixing (CSU) method is first evaluated on two synthetic images $I_1$ and $I_2$ of size $100 \times 100$ pixels and $L=224$ spectral bands. There are $R_0=5$ endmembers present in the images which correspond to the minerals Dipyre, Spodumene, Clinoptilolite, Mordenite and Olivine 1. Their spectral signatures have been obtained from the USGS spectral library \cite{Clark2007USGS} and are depicted in Fig. \ref{fig:endmembers_synth}. Note that the angles between the spectral signatures are between $3.01^\circ$ and $3.05^\circ$ (\emph{i.e.}, the endmembers are highly correlated with mutual coherence \cite{Donoho2006} that equals to $M=0.9986$), making the unmixing problem very challenging. The support maps that determine the spatial distribution of each material have been generated by simulating from the prior model \eqref{eq:MRF} with $\bbeta=[0.2;0.275;0.35;0.425;0.5]\transp$, and are depicted in the top row of Fig. \ref{fig:Potts_synth_R5}. For both images the matrix $\bfX$ has been generated by sampling from \eqref{eq:prior_x} with $s_r=0.3,\forall r$. Finally, the average signal-to-noise ratio (SNR) for the images $I_1$ and $I_2$ are approximately $30$dB ($\sigma_{\ell}^2=8.10^{-4}, \forall \ell$) and $20$dB ($\sigma_{\ell}^2=8.10^{-3}, \forall \ell$), respectively. 

\subsection{Supervised unmixing}
In this first experiment we consider that the materials present in the images $I_1$ and $I_2$ are perfectly known and we estimate the abundance vectors with CSU, CLSunSAL and NCLS. The CSU algorithm has been implemented using $N_{\textrm{MC}}=3000$ and $N_{\textrm{bi}}=1000$, and by allowing the algorithm to self-adjust the regularisation parameter $\bbeta$ with the technique proposed in \cite{Pereyra2014ssp}. The estimated values of $\bbeta$ are $\bbeta=[0.19,0.28,0.33,0.37,0.44]\transp$ and $\bbeta=[0.20,0.28,0.33,0.33,0.44]\transp$ for $I_1$ and $I_2$, respectively. For comparison we use the state-of-the-art sparse regression algorithm SunSAL \cite{Bioucas2010whispers} and the SunSAL algorithm using the total variation regularization (SunSAL-TV) \cite{Iordache2012}, whose regularisation parameter values are adjusted to provide the best abundance estimates (in terms of RMSE) for each scenario. For completeness we also compare with the widely used Non-negatively Constrained Least-Squares algorithm (NCLS) \cite{Heinz2001}, which solves a maximum likelihood problem and does not exploit any prior information about the abundance vectors.

The associated computation times for a Matlab implementation on a $3.0$GHz Intel Xeon quad-core workstation are provided in Table \ref{tab:time_synth1}. The estimated supports obtained with CSU, SunSAL, SunSAL-TV and NCLS for $I_2$ (lowest SNR) are depicted in Fig. \ref{fig:Potts_synth_R5}. The presence maps associated with $I_1$ are similar and are not presented here due to space constraints. For SunSAL, SunSAL-TV and NCLS, the detection maps have been obtained by thresholding the estimated abundances with a threshold arbitrarily set to $\rho=0.01$. We observe that the results obtained with CSU are in good agreement with the ground truths. On the other hand, the abundances obtained with SunSAL, SunSAL-TV and NCLS are significantly less accurate, thus confirming that taking into account the spatially correlations between the supports abundance vectors is key to achieving accurate estimation results in this scenario. This valuable prior knowledge is all the more important in low SNR scenarios such as the one depicted in Fig. \ref{fig:Potts_synth_R5}.

For numerical comparison, we computed the root mean square error (RMSE) 
$\textrm{RMSE}_{n}= \sqrt{\norm{\hat{\Vabonds}_{n} - \Vabond{n}}^2}$, 
that quantifies the average accuracy of the estimated abundances $\hat{\Vabonds}_{n}$ with respect to the truth $\Vabond{n}$ at the $n$-th pixel. 
We also consider the abundance angle distance (AAD) $\textrm{AAD}_{n}= \cos^{-1} \left(\dfrac{\hat{\Vabonds}_{n}\transp\Vabond{n}}{\norm{\hat{\Vabonds}_{n}}_2\norm{\Vabond{n}}}_2 \right)$,  
which is not sensitive to scaling factors between actual and estimated abundance vectors.
The lower the RMSEs and AADs, the better the abundance estimation performance.

The first five rows of Table \ref{tab:abund_synth1} show the average RMSEs and AADs,  for CSU, SunSAL, SunSAL-TV, NCLS, and the oracle NCLS (o-NCLS), which consists of applying NCLS using only the active materials in each pixel (\emph{i.e.}, with perfect knowledge of the support of the abundance vectors). We observe that CSU provides significantly more accurate estimations than SunSAL, SunSAL-TV and NCLS, achieving average RMSEs and AADs that are close to the oracle. Again, the good performance of CSU results from the collaboration between pixels introduced by the prior model \eqref{eq:MRF}. By comparing the first and second columns of Table \ref{tab:abund_synth1} we confirm that this prior knowledge becomes all the more important as the noise level increases. These results show that using only sparsity (SunSAL) does not lead to significant improvements, perhaps because the $\ell_1$ regularisation is not appropriate for our images, or because it would require adapting the regularisation parameters to each endmember (this is done automatically in CSU). We also observe that SunSAL-TV does not achieve significantly better results, perhaps due to the regions where the abundances exhibit fluctuations that are not well modelled by the TV. As mentioned previously, when both the abundance supports and the abundance values exhibit significant spatial dependencies, one should consider methods that account for this prior information (e.g., the SunSAL-TV method if the abundance maps are expected to be piecewise constant across the image). However, in many high-resolution images and regions in images, only the abundance supports exhibit significant spatial dependencies; in this case regularising the abundance values may lead to over-smooth estimation results. In this paper we show that by modelling these dependencies it is possible to improve unmixing performance with respect to methods that assume that abundances are uncorrelated.

\begin{table}[h!]
\renewcommand{\arraystretch}{1.2}
\begin{footnotesize}
\begin{center}
\caption{Computational time (in seconds): synthetic images
.\label{tab:time_synth1}}
\begin{tabular}{|c|c|c|c|}
\cline{3-4}
\multicolumn{2}{c|}{} &  $I_1$  & $I_2$  \\
\hline
\multicolumn{2}{|c|}{o-NCLS} & $5.08$ & $14.54$ \\
\hline
\multirow{3}*{R=5} & CSU & $3360$ & $3420$ \\
\cline{2-4}
									 & SunSAL & $0.62$ & $0.57$\\
\cline{2-4}
									 & SunSAL-TV & $56$ & $55$\\
\cline{2-4}
									 & NCLS & $0.67$ & $0.63$\\
\hline
\multirow{3}*{R=7} & CSU & $51330$ & $52500$\\
\cline{2-4}
									 & SunSAL & $0.55$ & $0.60$\\
\cline{2-4}
									 & CLSunSAL & $0.62$ & $0.57$\\
\cline{2-4}
									 & SunSAL-TV & $56$ & $53$\\
\cline{2-4}
									 & NCLS & $2.03$ & $3.40$\\
\hline
\end{tabular}
\end{center}
\end{footnotesize}
\vspace{-0.4cm}
\end{table}

\begin{table}[h!]
\renewcommand{\arraystretch}{1.2}
\begin{footnotesize}
\begin{center}
\caption{Average RMSEs and AADs: synthetic images
.\label{tab:abund_synth1}}
\begin{tabular}{|c|c|c|c|c|c|}
\cline{3-6}
\multicolumn{2}{c|}{} & \multicolumn{2}{|c|}{Av. RMSEs ($\times 10^{-2}$)} & \multicolumn{2}{|c|}{Av. AADs ($\times 10^{-2}$)}\\
\cline{3-6}
\multicolumn{2}{c|}{} &  $I_1$  & $I_2$ &  $I_1$  & $I_2$ \\
\hline
\multicolumn{2}{|c|}{o-NCLS} & $6.01$ & $17.20$ & $7.56$ & $20.91$\\
\hline
\multirow{3}*{R=5} & CSU & $6.30 $ & $17.05$ & $8.07 $ & $21.32$\\
\cline{2-6}
									 & SunSAL & $8.50$ & $23.52$ & $12.32$ & $32.35$\\
\cline{2-6}
									 & SunSAL-TV & $8.26$ & $22.96$ & $11.57$ & $31.55$\\
\cline{2-6}
									 & NCLS & $8.50$ & $23.54$ & $11.57$ & $31.52$\\
\hline
\multirow{3}*{R=7} & CSU & $6.61$ & $17.36$ & $8.44$ & $21.69$\\
\cline{2-6}
									 & SunSAL & $13.13$ & $31.87$ & $19.04$ & $43.87$\\
\cline{2-6}
									 & CLSunSAL & $13.11$ & $31.85$ & $18.73$ & $43.87$\\
\cline{2-6}
									 & SunSAL-TV & $13.06$ & $31.68$ & $18.01$ & $43.45$\\
\cline{2-6}
									 & NCLS & $13.13$ & $31.89$ & $18.01$ & $43.86$\\
\hline
\end{tabular}
\end{center}
\end{footnotesize}
\vspace{-0.4cm}
\end{table}

Finally, in order to further highlight the good performance of the proposed prior model we have also computed the reconstruction error
(RE) defined as $\textrm{RE}_{n}= \sqrt{\norm{\MATmat\hat{\Vabonds}_{n} - \Vpix{n}}^2}$.
Note that this error essentially measures the likelihood of the abundance estimates given the observed image $\MATpix$ and assuming the noise is identically distributed in the $L$ spectral bands. However, because the SU problem is not well-posed, the capacity of the likelihood to identify good solutions is severely limited and the additional information provided by the prior model is key to deliver accurate estimation results. In these scenarios Bayesian methods, which combine observed and prior information, can greatly outperform other estimation techniques. The average RE for each method is reported in Table \ref{tab:RE_synth1}. We observe that all the algorithms considered exhibit very similar REs. However, we know from Table \ref{tab:abund_synth1} that CSU outperforms significantly the other methods in terms of abundance estimation accuracy. The contrast between these two figures of merit confirms that the superior performance of CSU is directly related to the proposed prior model, which captures the correlations between the supports of the abundance vectors and effectively introduces a means of collaboration that allows sharing information between pixels and increasing robustness to noise.

\begin{table}[h!]
\renewcommand{\arraystretch}{1.2}
\begin{footnotesize}
\begin{center}
\caption{Average REs ($\times 10^{-2}$): synthetic images
.\label{tab:RE_synth1}}
\begin{tabular}{|c|c|c|c|}
\cline{3-4}
\multicolumn{2}{c|}{} &  $I_1$  & $I_2$  \\
\hline
\multirow{3}*{R=5} & CSU & $2.81$ & $8.89$\\
\cline{2-4}
									 & SunSAL & $2.80$ & $8.86$\\
\cline{2-4}
									 & SunSAL-TV & $2.80$ & $8.86$\\
\cline{2-4}
									 & NCLS & $2.80$ & $8.86$\\
\hline
\multirow{3}*{R=7} & CSU & $2.81$ & $8.89$\\
\cline{2-4}
									 & SunSAL & $2.80$ & $8.87$\\
\cline{2-4}
									 & CLSunSAL & $2.80$ & $8.87$\\
\cline{2-4}
									 & SunSAL-TV & $2.80$ & $8.86$\\
\cline{2-4}
									 & NCLS & $2.80$ & $8.86$\\
\hline
\end{tabular}
\end{center}
\end{footnotesize}
\vspace{-0.4cm}
\end{table}

\subsection{Semi-supervised unmixing}
We now consider that two additional endmembers are incorrectly included in the library $\MATmat$, although they are not present in the scene. To make the SU problem particularly challenging, the two additional endmembers are Olivine 2 and Adularia, whose spectral signatures are highly correlated with the signatures of the other endmembers. For example, note from Fig. \ref{fig:endmembers_synth} that discriminating between Olivine 1 and Olivine 2 is very difficult. We contrast CSU with SunSAL, SunSAL-TV and NCLS (CSU is implemented using $N_{\textrm{MC}}=7000$ and $N_{\textrm{bi}}=5000$ and the estimated values of $\bbeta$ are $\bbeta=[0.20,0.28,0.35,0.40,0.45,0.47,0.47]\transp$ and $\bbeta=[0.20,0.29,0.36,0.37,0.45,0.46,0.42]\transp$ for $I_1$ and $I_2$, respectively.). We also consider CLSunSAL \cite{Iordache2014a} which promotes group-sparsity for the estimated abundances.

\begin{figure*}[ht]
  \centering
    \includegraphics[width=0.9\textwidth]{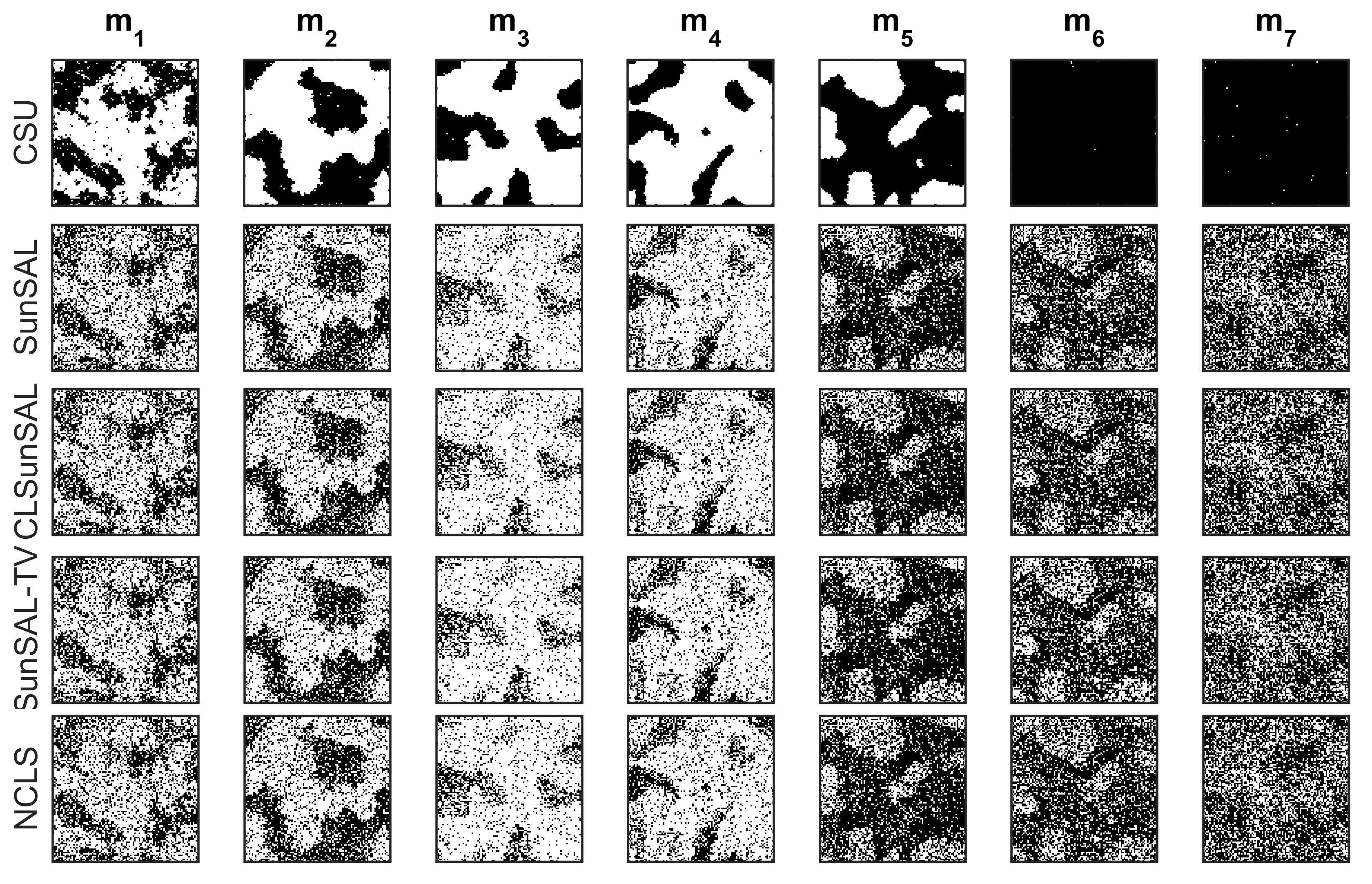}
  \caption{Active support maps estimated with the proposed CSU algorithm, SunSAL, CLSunSAL, SunSAL-TV and NCLS (top to bottom, $\rho=0.01$) for $I_2$ and $R=7$ endmembers. White (resp. black) pixels correspond to regions where a component is present (resp. absent).}
	\label{fig:Potts_synth_R7}
\end{figure*}

Fig. \ref{fig:Potts_synth_R7} shows the support estimations obtained with CSU, SunSAL, CLSunSAL and SunSAL-TV for $I_2$ (lowest SNR) with the $R=7$ endmembers (two of which are absent from the scene). The results obtained for $I_1$ follow the same trend and are not presented here. Again, we observe that the results obtained with CSU are in very good agreement with the ground truths, whereas the results obtained with the other methods are significantly less accurate, even when tuning the regularization parameter. More importantly, we observe that CSU has successfully detected that two endmembers (Olivine 2 and Adularia) are not present in the scene, in spite of the strong similarities between Olivine 1 and Olivine 2.

\begin{figure}[h!]
\centering
\includegraphics[width=\columnwidth]{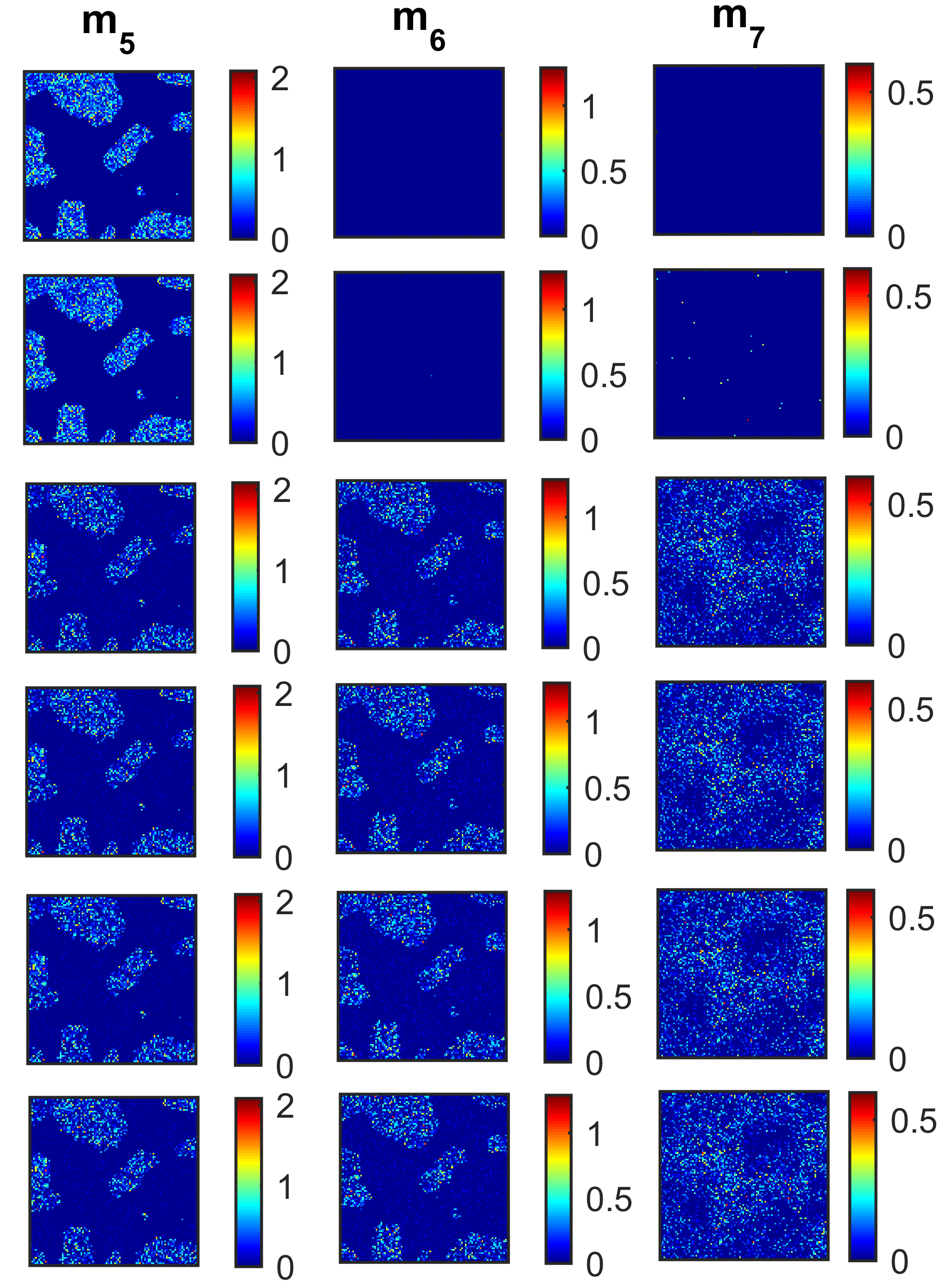}
\caption{Top: Actual abundance maps of the endmembers $5, 6$ and $7$ for $I_2$. Abundance maps estimated by CSU, SunSAL, CLSunSAL, SunSAL-TV and NCLS (top to bottom).}
\label{fig:Abund_synth_R7_SNR2}
\end{figure}

Finally, the five bottom rows of Table \ref{tab:abund_synth1} show the RMSEs and AADs obtained with CSU, SunSAL, SunSAL-TV, CLSunSAL and NCLS for $I_1$ and $I_2$ with $R=7$. We observe that CSU is robust to the presence of the two redundant endmembers and is able to accurately discriminate between Olivine 1 and Olivine 2. On the other hand, SunSAL, SunSAL-TV, CLSunSAL and NCLS have difficulties detecting the true supports of the abundance vectors and produce estimation results that are significantly less accurate, in particular in low SNR conditions. Again, the superior performance of CSU is directly related to the prior model \eqref{eq:MRF} that captures the spatial correlations of the mixture support in the image and regularises the supports of the abundance vectors. Notice that the group-sparsity model operating in CLSunSAL is unable to identify correctly the endmembers present in the images (due to the high correlation between the two Olivine spectra). Increasing the CLSunSAL regularisation parameter leads to a correct identification of the endmembers, but at the expense of a severe degradation in estimation performance. Similarly, increasing the SunSAL-TV regularisation parameter improves the detection of absent endmembers, but degrades overall estimation performance. Notice that CSU does not suffer from these drawbacks because it self-adjusts the scale parameters ($\bfs^2$) and spatial correlation parameters ($\bbeta$) automatically during the inference procedure. For completeness, Fig. \ref{fig:Abund_synth_R7_SNR2} presents the abundances of the endmembers $5,6$ and $7$ (which are the hardest to discriminate) estimated with the different algorithms for $I_2$. 
\section{Application to real a hyperspectral image}
\label{sec:simu_real}

\begin{figure}[h!]
  \centering
  \includegraphics[width=\columnwidth]{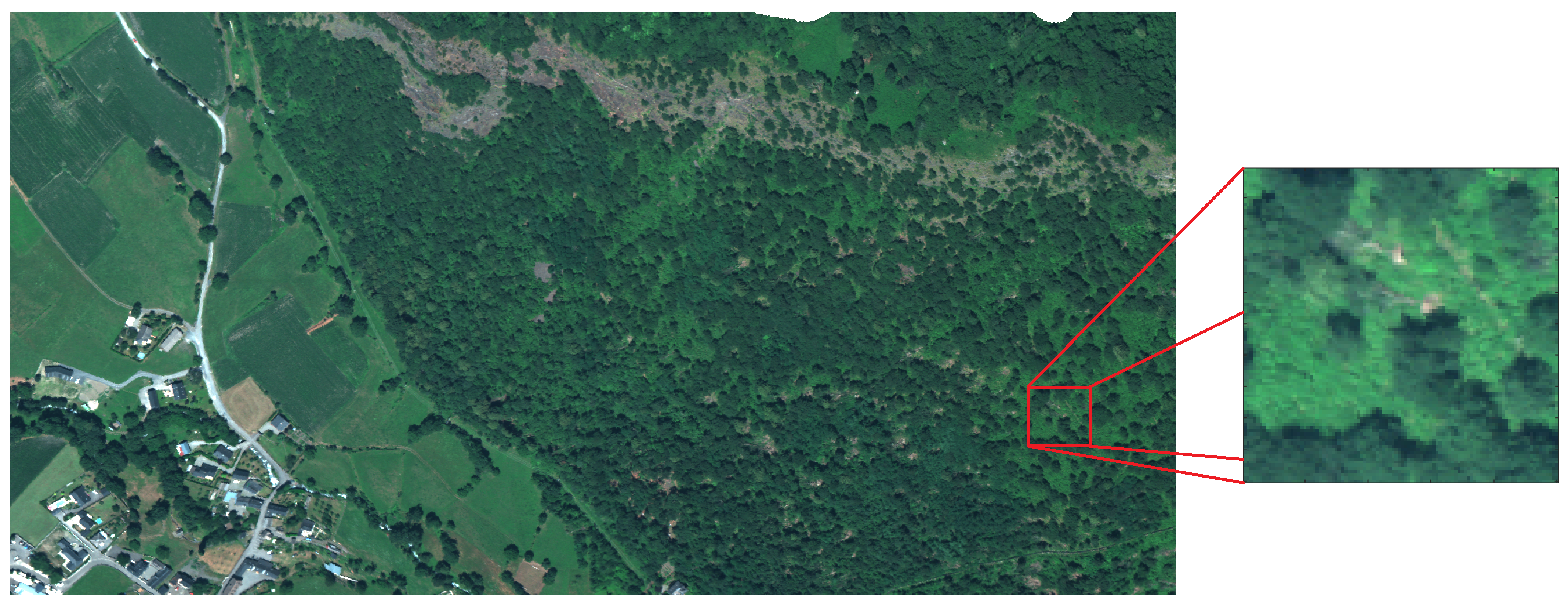}
  \caption{True color image of the Villelongue area (left) and sub-image of interest (right).}
  \label{fig:Madonna_big}
\end{figure}

This section presents an application of the proposed CSU method to a real hyperspectral image acquired by the Hyspex hyperspectral scanner over Villelongue,
France ($00^\circ03'$W and $42^\circ57'$N). This images was acquired in 2010 as part of the Madonna project, and is composed of $L=
160$ spectral bands covering from the visible to near infrared spectrum and with a spatial resolution of $0.5$m (for more details about the data acquisition and pre-processing steps see \cite{Sheeren2011}.). This dataset has previously been studied
in \cite{Sheeren2011,Altmann2014a,Altmann2013} and is mainly composed of forested and urban areas. Due to the high spatial resolution of the image, materials present in a pixel are likely to be also present in its neighbour pixels; that is, we expect significant spatial correlations between the mixture supports.  Here we evaluate the proposed unmixing method on the region of interest of size $100 \times 100$ pixels depicted in Fig. \ref{fig:Madonna_big}. This region is composed mainly of trees and grass, with $R=6$ endmembers related to soil, two types of grass, two types of trees, and an additional endmember modelling attenuation effects mainly related to shade. The spectral signatures for these endmembers have been extracted manually from the data by using our prior knowledge about the scene \cite{Sheeren2011}. Note that for this image and within the LMM framework, the attenuation effects have to be modelled as an additional endmember because they operate differently on the different spectral bands of the reflectance spectra (otherwise, if the attenuations acted as a scaling factor affecting all the bands similarly, we would be able to handle this by relaxing the abundance sum-to-one constraint).

\begin{figure}[h!]
\centering
\includegraphics[width=\columnwidth]{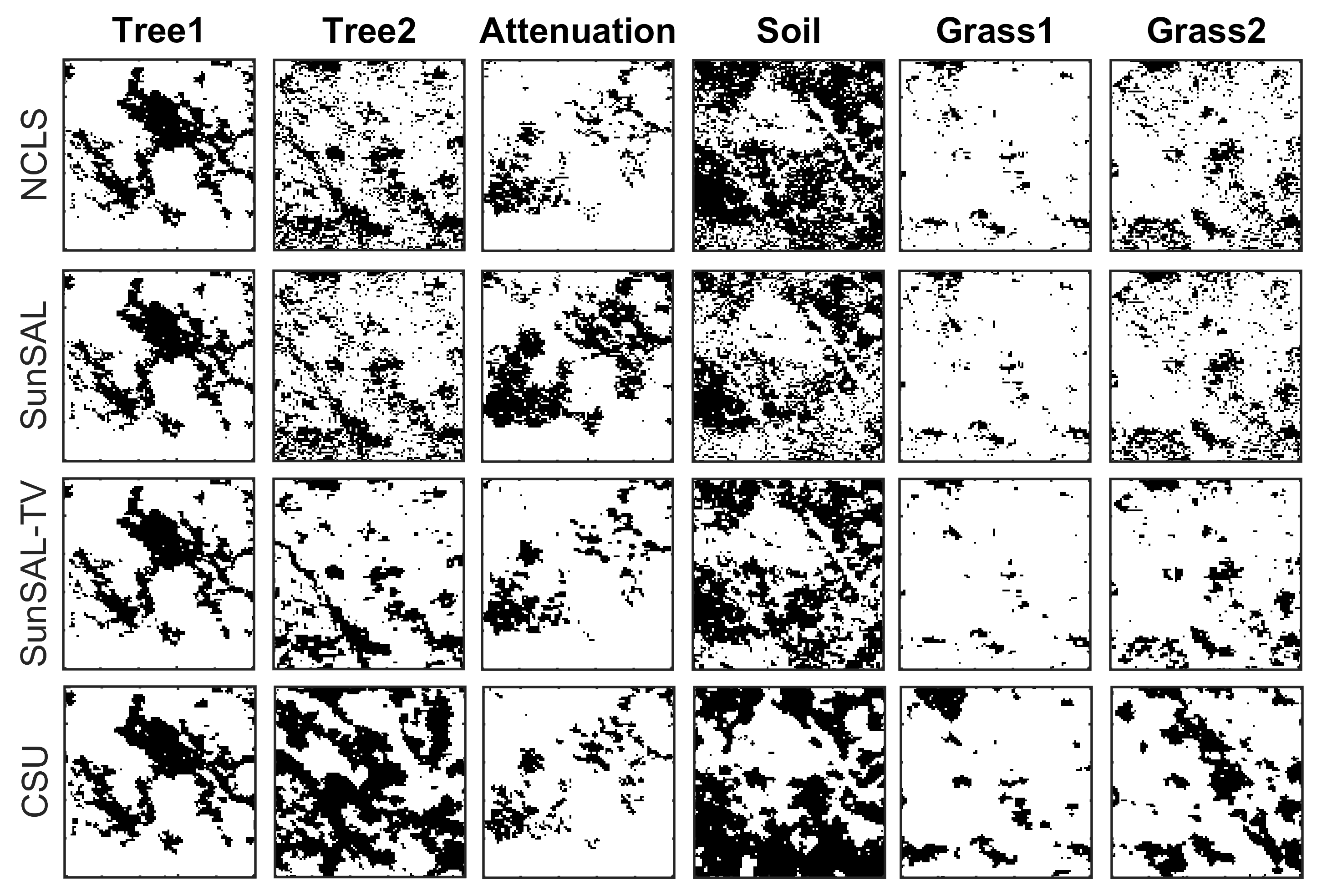}
\caption{Estimated presence maps (white pixels correspond to regions where the endmembers are present) for the real Villelongue image.}
\label{fig:detection_Madonna}
\end{figure}

\begin{figure}[h!]
\centering
\includegraphics[width=\columnwidth]{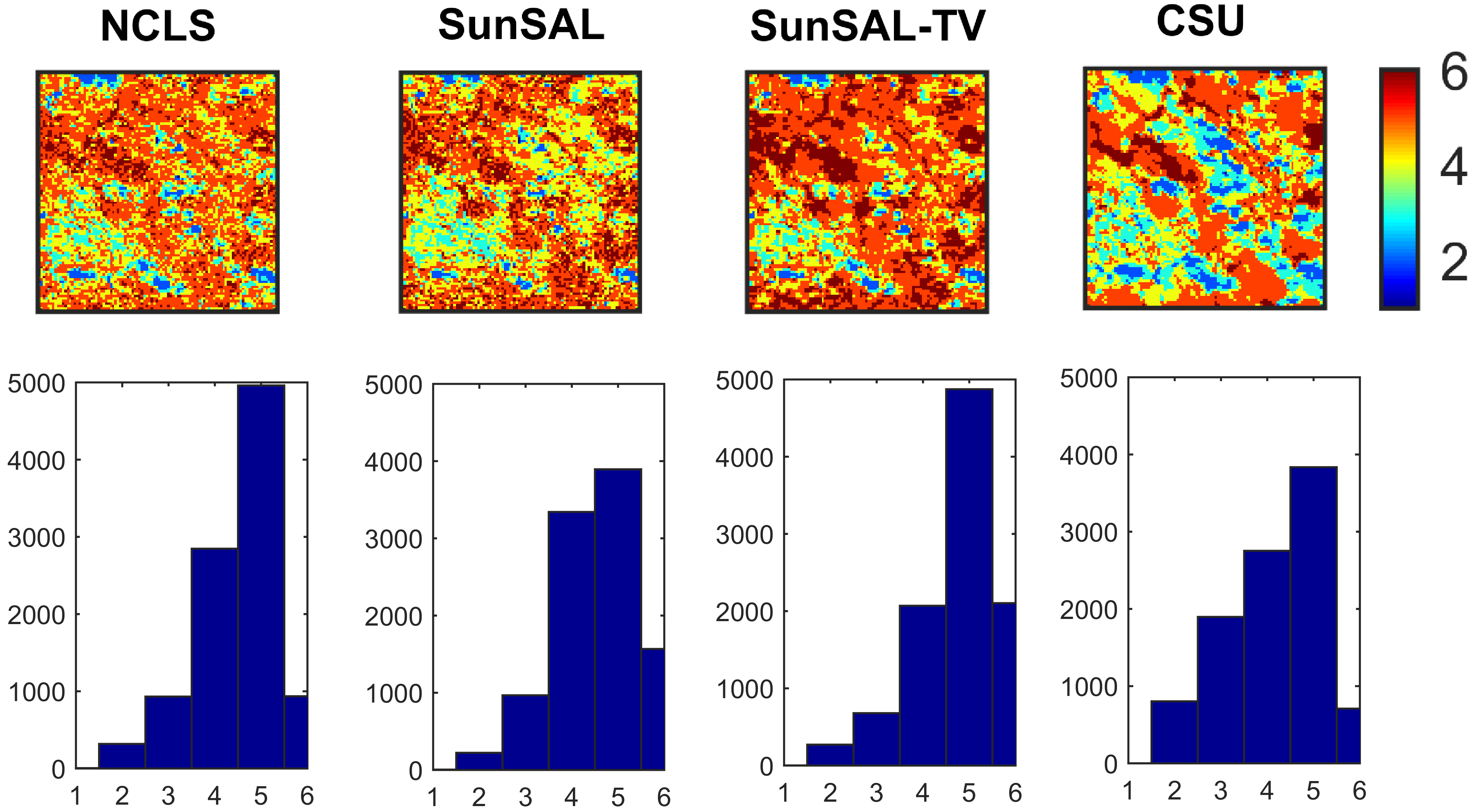}
\caption{Top: Estimated number of active endmembers per pixel in the Villelongue scene. Bottom: Histograms of the estimated numbers of active endmembers per pixel (computed over the $10^{5}$ image pixels)}
\label{fig:Supports_Madonna}
\end{figure}

\begin{figure*}[ht]
\centering
\includegraphics[width=0.9\textwidth]{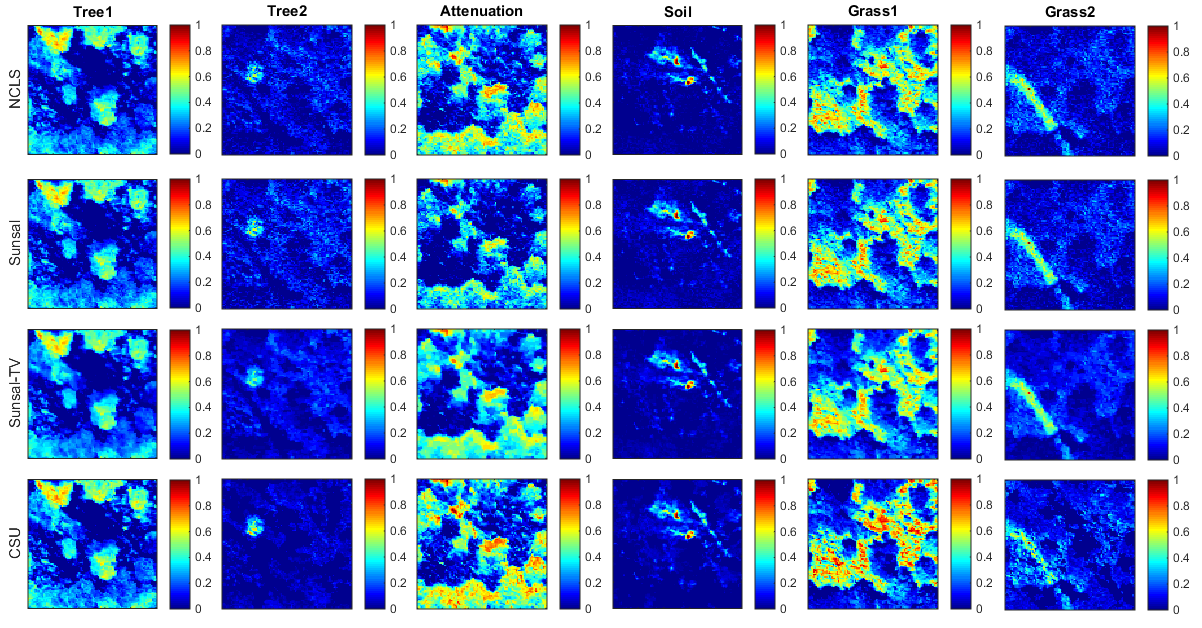}
\caption{Estimated abundance maps for the real Villelongue image.}
\label{fig:estimation_Madonna}
\end{figure*}

Fig. \ref{fig:detection_Madonna} shows the presence maps for each material estimated with CSU (using $N_{\textrm{MC}}=5000$, $N_{\textrm{bi}}=1000$ and by allowing the algorithm to self-adjust the regularisation parameter $\bbeta$ with the technique \cite{Pereyra2014ssp}) and with NCLS, SunSAL and SunSAL-TV (whose parameters have been optimised in a fully supervised manner to obtain the sparser and/or smoother abundance maps without degrading significantly the reconstruction error, and by using a detection threshold of $\rho=0.01$). Fig. \ref{fig:Supports_Madonna} shows the total number of materials present in each pixels, as estimated by each method. Lastly, Fig. \ref{fig:estimation_Madonna} shows the abundance maps estimated with each method. These figures clearly show that CSU provides sparser and spatially smoother presence (support) maps than the other methods. We also observe that, as expected, SunSAL-TV provides smoother abundance maps than the other methods (see Fig. \ref{fig:estimation_Madonna}). This property of SunSAL-TV is beneficial in regions where the materials are absent; however, where materials are present, it may smooth out fine detail (e.g., textures). Since no abundance ground truth is available for this data set, it is difficult to determine which algorithm provides the more accurate results. However, due to the high spatial resolution of the image and the structure of the materials considered, it is reasonable to assume that abundances will exhibit some degree of local heterogeneity, and that CSU and SunSAL-TV will outperform each other in different ways. Moreover, from Figs. \ref{fig:detection_Madonna} and Fig. \ref{fig:estimation_Madonna} we also note that CSU detects a higher lever of attenuation effects than SunSAL and SunSAL-TV, particularly in pixels containing vegetation where strong shadowing effects occur.

\begin{figure}[h!]
  \centering
  \includegraphics[width=\columnwidth]{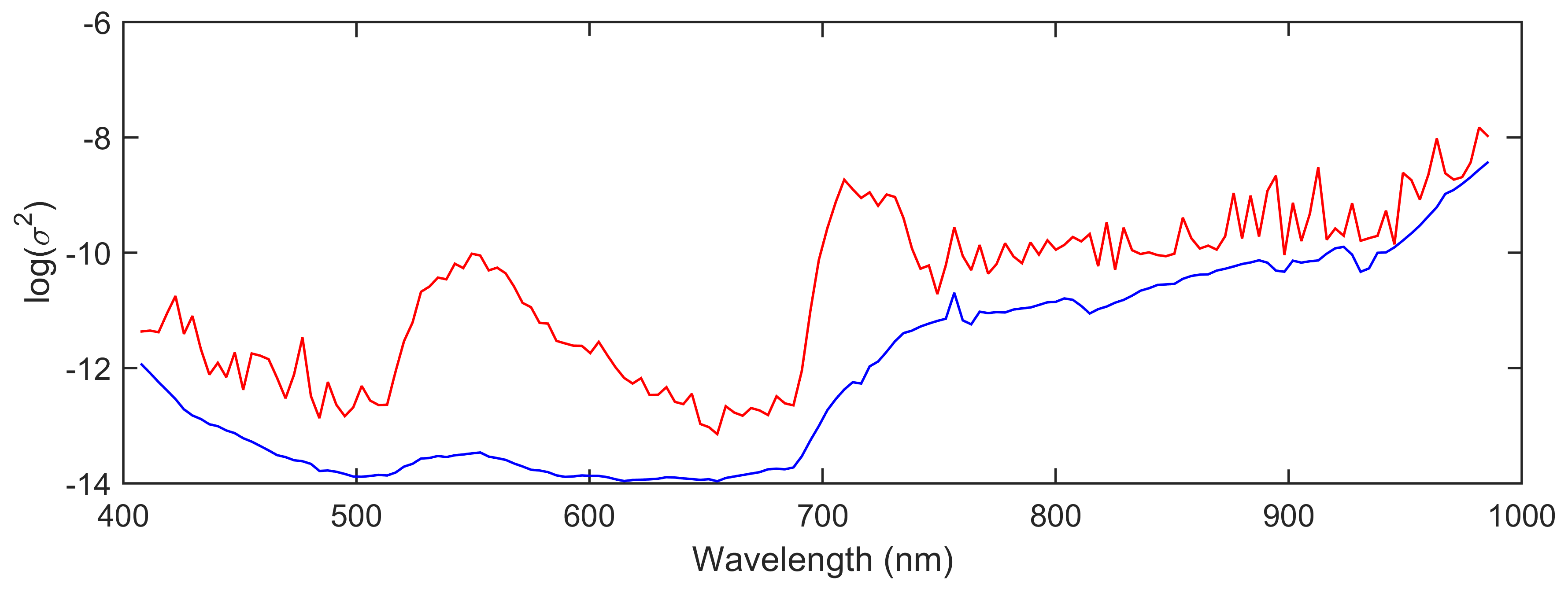}
  \caption{Noise variances estimated by the CSU (blue) and the Hysime
algorithms (red) for the real Villelongue image.}
  \label{fig:noise_Madonna}
\end{figure}

Finally, as a model checking analysis, we plot in Fig. \ref{fig:noise_Madonna} the marginal noise variances estimated by CSU (i.e., diagonal elements of the full estimated covariance matrix).  For comparison we also include the estimates obtained with Hysime \cite{Bioucas2008}. We observe that the estimation results produced by CSU are very smooth and physically realistic, suggesting good fit to data (recall that CSU assumes that the noise variances are prior independent, hence the smoothness of these posterior estimates arises from the data and from correlations with other model parameters). We also note that, although the estimation results produced by CSU and Hysime are different, the noise variances follow similar profiles. Lastly, for completeness Table \ref{tab:RE_real} reports the reconstruction errors associated with CSU, NCLS, SunSAL and SunSAL-TV. We observe that all methods achieve comparable reconstruction errors, with NCLS and SunSAL having a slightly lower reconstruction error than CSU and SunSAL-TV. This difference is likely due to the fact that NCLS and SunSAL do not enforce spatial regularity, and are hence more prone to overfitting the data.

\begin{table}[h!]
\renewcommand{\arraystretch}{1.2}
\begin{footnotesize}
\begin{center}
\caption{Average reconstruction errors (ARE): Real image.\label{tab:RE_real}}
\begin{tabular}{|c|c|}
\hline
\multirow{1}*{Unmixing algo.} &  ARE ($\times 10^{-2}$) \\
\hline
\multicolumn{1}{|c|}{CSU} & $7.14$ \\
\hline
\multicolumn{1}{|c|}{SunSAL} & $6.98$ \\
\hline
\multicolumn{1}{|c|}{SunSAL-TV} & $7.00$ \\
\hline
\multicolumn{1}{|c|}{NCLS} & $6.88$ \\

\hline
\end{tabular}
\end{center}
\end{footnotesize}
\vspace{-0.4cm}
\end{table}

\section{Conclusion}\label{sec:conclusion} 
This paper presented a new Bayesian method for linear unmixing of hyperspectral image that is based on a collaborative sparse regression formulation. The main novelty is a Bayesian model for sparse regression that takes into account the fact that the supports of the abundance vectors are spatially correlated. This prior information is encoded in the model by using a truncated multivariate Ising Markov random field model, which also accounts for the facts that pixels cannot be empty (\emph{i.e.}, there is at least one material per pixel) and that each material in the scene may require a different amount of spatial regularisation. The proposed Bayesian model also takes into consideration that material abundances are non-negative quantities and that the level of noise contaminating the image may be unknown. Following on from this, we presented a Markov chain Monte Carlo algorithms to perform Bayesian inference with this model and compute the statistical estimators of interest. Precisely, we proposed a Gibbs sampler that allows estimating the probabilities that materials are present or absence in each pixel, and, conditionally on any given configuration (typically the maximum \emph{a posteriori}), computing the MMSE estimates of the abundance vectors. A remarkable characteristic of the proposed inference algorithm is that it self-adjusts the amount of regularity enforced by the random field, thus relieving practitioners from the difficult task of setting regularisation parameters by cross-validations. Finally, the good performance of the proposed methodology was demonstrated through a series of experiments with synthetic and real data and comparisons with other algorithms from the literature. 

Due to its computational complexity, the algorithm presented in this paper can only be directly applied to problems with small numbers of endmembers (\emph{e.g.}, $R \leq 25$). For problems with larger libraries it is computationally more efficient to generate samples from a relaxed posterior in which the ``non-empty pixel'' constraint \eqref{eq:psi} of the MRF is removed, and then reintroduce this constraint by importance sampling. Another possibility for problems with large libraries is to use the MUSIC-CSR algorithm \cite{Iordache2014b} as a pre-processing step to identify endmembers that are absent from the scene, and then apply our method using a pruned library.

As mentioned previously, previous works have considered that the values of the abundance vectors are spatially correlated, whereas we considered correlations between their supports (\emph{i.e.}, material presence and absence patterns). Investigating new models that exploit both types of correlations is an important perspective for future work. For many applications or hyperspectral datasets, it makes sense to consider additional abundance constraints, such as the sum-to-one constraint to further improve unmixing results. Embedding this constraint within $\ell_0$-type sparse regression models is a challenging problem that is currently under investigation. Another perspective for future work is to investigate more sophisticated spatial models that describe hyperspectral images more accurately, in particular in complex scenes with numerous materials and non-linear effects. Finally, we believe that the methodology presented in this paper could be interesting for other regression problems that exhibit structured sparsity and intend to further investigate this is the future.

\section*{Acknowledgments}
The authors would like to thank Prof. Jean-Yves Tourneret and Dr Nicolas Dobigeon, from the University of Toulouse, IRIT-ENSEEIHT, France, for interesting discussion regarding this work. 

\section*{Appendix: On the conditional distribution of the abundance values}
Due to the conjugacy of the priors (9) and the likelihood (3), the posterior distribution 
\begin{eqnarray}
f(\bfx_{n}|\Vpix{n},\bfz_{n},\bsigma^{2},\bs^{2}) \propto f(\Vpix{n}|\bfx_{n},\bfz_{n},\bsigma^{2}) f(\bfx_{n}|\bs^{2})\nonumber
\end{eqnarray}
is a multivariate Gaussian distribution restricted to $\Omega$, i.e., the positive orthant of $\bbR^{R}$. 
Moreover,$\forall \bfx_{n} \in \Omega$\\
$\log\left( f(\bfx_{n}|\Vpix{n},\bfz_{n},\sigma^{2},\bs^{2})\right)$ 
\begin{eqnarray}
 & =  \log\left(f(\Vpix{n}|\bfx_{n},\bfz_{n},\bsigma^{2})\right) \log\left(f(\bfx_{n}|\bs^{2})\right) ,\nonumber\\
 & =  -\dfrac{\tilde{\mathbf{y}}_n\transp\bSigma_0^{-1}\tilde{\mathbf{y}}_n}{2} -\dfrac{\bfx_{n}\transp\bfS^{-1}\bfx_{n}}{2}+c_1\nonumber\\
 & =  -\dfrac{1}{2} \bfx_{n}\transp\left(\bfS^{-1} + \bfD_{n}\MATmat\transp\bSigma_0^{-1}\MATmat\bfD_{n} \right)\bfx_{n}\nonumber\\
 & +  \bfx_{n}\transp\bfD_{n}\MATmat\transp\bSigma_0^{-1}\Vpix{n} + c_2\nonumber
\end{eqnarray}
where $\bfS=\textrm{diag}(\bfs^2)$ and $\bfD_{n}=\textrm{diag}(\bfz_{n})$ are diagonal matrices with diagonal elements given by $\bfs^2$ and $\bfz_{n}$ and $c_1,c_2$ are real constants (independent of $\bfx_{n}$).
By identification, we obtain
$\bSigma_{n}^{-1}  =  \bfD_{n}\MATmat\transp\bSigma_0^{-1}\MATmat\bfD_{n} + \bfS^{-1}$, $\bSigma_{n}^{-1}\bar{\bfx}_{n}  =  \bfD_{n}\MATmat\transp\bSigma_0^{-1}\Vpix{n}$
and finally 
\begin{eqnarray}
\bSigma_{n} & = & \left(\bfD_{n}\MATmat\transp\bSigma_0^{-1}\MATmat\bfD_{n} + \bfS^{-1}\right)^{-1}\nonumber\\
\bar{\bfx}_{n} & = & \bSigma_{n} \bfD_{n}\MATmat\transp\bSigma_0^{-1}\Vpix{n},\nonumber
\end{eqnarray}

\clearpage
\bibliographystyle{IEEEtran}
\bibliography{biblio}

\end{document}